# Effect of the steady flow on spatial damping of small-amplitude prominence oscillations

K.A.P Singh[1], R. Erdélyi [2] and B.N. Dwivedi [3]

1. Kodaikanal Solar Observatory, Indian Institute of Astrophysics, Kodaikanal-624103, India.
2. Solar Physics and Space Plasma Research Centre (SP$^2$RC), Department of Mathematics, University of Sheffield, Hicks Building, Hounsfield Road, Sheffield S3 7RH, UK.
3. Department of Applied Physics, Institute of Technology, Banaras Hindu University, Varanasi-221005, India.

e-mails: alkendra@iiap.res.in; robertus@sheffield.ac.uk; bholadwivedi@gmail.com

## ABSTRACT

*Aims.* Taking account of steady flow in solar prominences, we study its effects on spatial damping of small-amplitude non-adiabatic magnetoacoustic waves in a homogeneous, isothermal, and unbounded prominence plasma.

*Methods.* We model the typical feature of observed damped oscillatory motion in prominences, removing the adiabaticity assumption through thermal conduction, radiation and heating. Invoking steady flow in MHD equations, we linearise them under small-amplitude approximation and obtain a new general dispersion relation for linear non-adiabatic magnetoacoustic waves in prominences

*Results.* The presence of steady flow breaks the symmetry of forward and backward propagating MHD wave modes in prominences. The steady flow has dramatic influence on the propagation and damping of magnetoacoustic and thermal waves. Depending upon the direction and strength of flow the magnetoacoustic and thermal modes can show both the features of wave amplification and damping. At the wave period of 5 min where the photospheric power is maximum, the slow mode shows wave amplification. However, in the absence of steady flow the slow mode wave shows damping.

*Conclusions.* For the wave period between 5 min and 15 min, the amplification length for slow mode, in the case of prominence regime 1.1, varies between $3.4 \times 10^{11}$ m to $2 \times 10^{12}$ m. Dramatic influence of steady flow on small-amplitude prominence oscillations is likely to play an important role in both wave detection and prominence seismology.

## 1. Introduction

Observations of oscillations in coronal structures hold the key to understanding the mechanisms for coronal heating and the solar wind acceleration. This also provides a unique opportunity to determine the coronal magnetic field and transport coefficients through *MHD seismology* (or *atmospheric magneto-seismology*). Coronal oscillations have been detected in several coronal structures, e.g. in *coronal loops* (Nakariakov et al. 1999; Aschwanden et al. 1999; De Moortel, Ireland & Walsh 2000; Robbrecht et al. 2001), in *coronal holes* (Ofman et al. 2000; Banerjee et al. 2001), in *coronal plumes* (De Forest & Gurman 1998; Banerjee, O'Shea & Doyle 2000) and in *prominences* (Terradas et al. 2002; Jing et al. 2003, 2006).

Observations from both ground and space show the presence of flows in various magnetic structures of the solar atmosphere. MHD waves in these structures may be influenced, or even generated by flows. One of the common influences of flows is the Doppler shift of the phase speed of waves, while noting that the case of shear flows at boundary layers can excite MHD waves due to instabilities. Moreover, the flow effects on waves can be steady as well as dynamic, depending on the ratio of characteristic time of waves propagating through the medium to characteristic time needed to shuffle the flow streamlines. Even if the flow streamlines are slightly re-organized during the time a wave passes through the domain of interest, one may consider steady flows (Terra-Homem, Erdélyi & Ballai 2003). Steady flows break the symmetry of forward (field-aligned) and backward (anti-parallel to magnetic field) propagating MHD wave modes because of induced Doppler shifts. Terra-Homem, Erdélyi & Ballai (2003) have shown that strong flows not only change the sense of propagation of MHD waves but also induce shifts in cut-off values and phase speeds of waves. In this work, we study the effect of steady flow on MHD wave propagation and damping in prominences when the changes in characteristic time scales are much longer than characteristic periods of waves.

Analysis of mass motions from center to limb observations shows the existence of a horizontal flow in and around prominence, the direction of the velocity being the same as that of magnetic field lines. The magnitude of these motions varies from 5 km s$^{-1}$ to 30 km s$^{-1}$ (cf., Schmieder 1989 for further details). High-resolution Hα images of prominences show the presence of numerous thin, thread-like magnetic structures and continuous plasma flow along these fibrils with velocities of 15 km s$^{-1}$ (Lin et al. 2005), which is suggestive of field-aligned flows. Observations of quiescent filaments in Hα yield velocities in the 5 km s$^{-1}$ to 20 km s$^{-1}$ range including the *counterstreaming* field-aligned flows (Zirker, Engvold & Martin 1998; Lin et al. 2003). SOHO observations yield horizontal flows of the order 10 km s$^{-1}$ along the line of sight (SUMER observations by Madjarska et al. 1999) and flow velocities in the range 2-13 km s$^{-1}$ in prominences (Patsourakos & Vial, 2002). Higher flow velocities have also been observed by TRACE in 'activated' prominence (Kucera, Tovar & de Pontieu 2003). In some cases it has been shown that the propagating waves move in the direction of the mass flows (Lin et al. 2007).

The temporal and spatial variations of prominence oscillations not only suggest their strong temporal damping but also the decrease of the amplitude of the propagating slow mode wave within a distance 2-5 × 10$^4$ km from the place of its generation (Terradas et al. 2002). Observations and theory of prominence oscillations and their damping have been widely reported (e.g., Ramsey & Smith 1966; Hyder, 1966; Bashkirtsev & Mashnich 1984; Balthasar et al. 1993; Molowny-Horas et al. 1999; Terradas, Oliver & Ballester 2001; Oliver & Ballester 2002; Ballai 2003; Engvold 2004; Carbonell, Oliver & Ballester 2004; Foullon, Verwichte & Nakariakov 2004; Terradas et al. 2005; Ballester 2006; Carbonell et al. 2006; Lin et al. 2007; Forteza et al. 2007; Soler, Oliver & Ballester 2007; Singh 2007; Singh, Dwivedi & Hasan 2007). In this paper, we consider this distance as the typical spatial damping length $d_L$ (= $\frac{1}{Im(k)}$) of the prominence oscillations. The adiabaticity assumption is removed through thermal conduction, radiation and heating. The resulting modes will have complex frequencies and perturbations will contain a spatially decaying exponential term. We include the effect of steady flow on the spatial damping of slow mode wave in prominences. To model the spatial damping of prominence oscillations, a homogeneous and unbounded equilibrium configuration is considered which is permeated by a magnetic field in x- direction. In sect. 2, we consider MHD equations invoking the effect of the steady flow to derive a new general dispersion relation for the linear and non-adiabatic magnetoacoustic wave propagation in solar prominences. Results and discussion are presented in sect. 3 followed by conclusions in the last section.

**2. Basic MHD equations**

We consider a homogeneous equilibrium configuration which is unbounded in all directions with magnetic field in the x-direction and neglect the effect of gravity. We have

$p_0$ = constant, $\rho_0$ = constant, $T_0$ = constant, and $\boldsymbol{B_0} = B_0\, \boldsymbol{e_x}$

where $p_0$, $\rho_0$, $T_0$ and $B_0$ respectively are equilibrium values of pressure, density, temperature and magnetic field. The basic equations for the discussion of linear and non-adiabatic magnetoacoustic waves are considered (cf. Carbonell et al. 2006). The non-adiabatic effects in the energy equation are invoked through a thermal conduction term $\nabla \cdot (\kappa \nabla T)$. Heat-loss function $L$ which depends on physical parameters, represents the difference between a radiative loss function and an arbitrary heat input given by $L(\rho, T) = \chi^* \rho\, T^\alpha - h\, \rho^a\, T^b$. The zeroth-order term in the radiative loss is taken as $L(\rho_0, T_0) = 0$. The optically thin and thick radiation are mimicked by choosing appropriate $\chi^*$ values and different heating scenarios are included corresponding to various combinations of the exponents $a$ and $b$. The prominence plasma can be optically thick in some spectral lines, so the radiative losses will be reduced and this can be modeled by changing the exponent α in the cooling function.

*2.1 General dispersion relation*

We consider small perturbations from the equilibrium in pressure, density, temperature, magnetic field and velocity of the form

**B** = **B**$_0$ + **B**$_1$ (*t*, *r*), $\rho = \rho_0 + \rho_1$ (*t*, *r*), $T = T_0 + T_1$ (*t*, *r*), $p = p_0 + p_1$ (*t*, *r*), **V**$_1$ = **V**$_0$ + **V**$_1$ (*t*, *r*).

We take perturbations proportional to *exp* i($\omega t$ - **k·r**) where **k** = $k_x$ **x** + $k_z$ **z** and $k_y = 0$. The linearised equations in terms of the Doppler-shifted frequency $\Omega = \omega - k_x V_0$ are given by:

$$\Omega \rho_1 - \rho_0 (k_x V_x + k_z V_z) = 0, \tag{1}$$

$$\Omega \rho_0 V_x = k_x p_1, \tag{2}$$

$$\Omega \rho_0 V_z = k_z p_1 + (k^2 B_0^2 V_z) / \mu (\Omega). \tag{3}$$

$$\Omega (p_1 - C_s^2 \rho_1) = i (A T_1 + H \rho_1), \tag{4}$$

$$\frac{p_1}{p_0} = \frac{\rho_1}{\rho_0} + \frac{T_1}{T_0}, \tag{5}$$

where A = ($\gamma$-1) ($k_{\parallel} k_x^2 + \rho_0 L_T$), H = ($\gamma$-1) (*L* + $\rho_0 L_\rho$), $L_\rho$ and $L_T$ are given by:

$$L_\rho = (\frac{\partial L}{\partial \rho})_T,$$

$$L_T = (\frac{\partial L}{\partial T})_p,$$

with *T* and *p* kept constant, respectively, at the equilibrium state and $C_s = \sqrt{\frac{\gamma RT}{\mu}}$ is the sound speed.

In the presence of a steady flow **V**$_0$, the Eulerian perturbation of **V**$_1$ and the Lagrangian displacement $\xi$, are related as **V**$_1$ = $\frac{\partial \xi}{\partial t}$ + $\nabla \times (\xi \times$ **V**$_0$) $- \xi \nabla \cdot$ **V**$_0$ + **V**$_0$ $\nabla \cdot \xi$. (cf. Erdélyi, Goossens & Ruderman 1995).

We obtain a general dispersion relation of the form,

$$a_6 k^6 + a_5 k^5 + a_4 k^4 + a_3 k^3 + a_2 k^2 + a_1 k + a_0 = 0 \tag{6}$$

where the coefficients are:

$$a_0 = \omega^5 - C_2 \omega^4,$$

$$a_1 = -5\omega^4 V_0 \cos\theta + 4C_2 V_0 \cos\theta\, \omega^3,$$

$$a_2 = 10 V_0^2 \cos^2\theta\, \omega^3 - (V_A^2 + C_s^2)\omega^3 - C_1\omega^4 + C_2\omega^2\left(-6V_0^2\cos^2\theta + V_A^2 + \frac{C_s^2}{\gamma}\right) - iH\omega^2$$

$$a_3 = -10 V_0^3 \cos^3\theta\,\omega^2 + 3V_0 \cos\theta\,\omega^2 (V_A^2 + C_s^2) + 4C_1 V_0 \cos\theta\,\omega^3 + C_2\omega\left(4V_0^3 \cos^3\theta\right.$$
$$\left. - 2V_A^2 V_0 \cos\theta - 2V_0 \frac{C_s^2}{\gamma}\cos\theta\right) + 2iH V_0 \cos\theta\,\omega$$

$$a_4 = 5V_0^4 \cos^4\theta\,\omega - 3(C_s^2 + V_A^2)V_0^2 \cos^2\theta\,\omega + C_s^2 V_A^2 \cos^2\theta\,\omega$$
$$- C_2 \cos^2\theta\left(V_A^2 \frac{C_s^2}{\gamma} + V_0^4 \cos^2\theta\right) + i(V_A^2 - V_0^2)H\cos^2\theta + C_1\left(V_A^2 + \frac{C_s^2}{\gamma}\right)\omega^2$$
$$- 6C_1 V_0^2 \cos^2\theta\,\omega^2 + C_2 V_0^2 \cos^2\theta\left(V_A^2 + \frac{C_s^2}{\gamma}\right)$$

$$a_5 = (V_A^2 + C_s^2)V_0^3 \cos^3\theta - V_A^2 C_s^2 V_0 \cos^3\theta - V_0^5 \cos^5\theta + \omega C_1\left(4V_0^3 \cos^3\theta - 2V_A^2 V_0 \cos\theta\right.$$
$$\left. - 2\frac{C_s^2}{\gamma}V_0 \cos\theta\right)$$

$$a_6 = C_1\left(\frac{C_s^2}{\gamma} - V_0^2 \cos^2\theta\right)(V_A^2 - V_0^2)\cos^2\theta$$

where $C_1 = \left(\dfrac{i 2 T_0 k_\parallel}{3 p_0}\right)\cos^2\theta$ and $C_2 = \dfrac{2}{3}\left(\dfrac{i T_0 \rho_0 L_T}{p_0}\right)$.

In this work we focus on the features of small-amplitude magnetoacoustic waves in prominence plasma invoking steady flow in MHD equations. The parameter values akin to various prominence regimes are given in Table 1. The dispersion relation (6) has been solved numerically for a magnetic field of 10 G and a propagation angle of π/4. Since we are interested in the spatial damping, we take $\omega$ to be real and wavenumber $k = k_R + i\, k_I$ to be complex. The behaviour of damping length as a function of steady flow is studied in the range of -15 km s$^{-1}$ to 15 km s$^{-1}$. It can be seen from our dispersion relation (6) that the static dispersion relation of Carbonell et al. (2006) can be retrieved by removing the flow terms. We have studied the wavelength $\lambda$ ($= \frac{2\pi}{k_R}$), the damping (or amplification) length $d_L$ ($= \frac{1}{Im(k)}$) and the damping per wavelength $D_L$ ($= \frac{k_I}{k_R}$) for different cases.

## 3. Results and discussion

*3.1 Behaviour of wave modes*

The inclusion of steady flow changes the propagation properties of slow, fast and thermal modes for the flow strength between -15 km s$^{-1}$ to 15 km s$^{-1}$. Steady flow breaks the symmetry of forward and backward propagating MHD waves. In case of prominence regime 1.1, it is found that the forward propagating slow and thermal wave modes show amplification whereas the fast mode wave shows damping when the flow strength is kept anti-parallel at 15 km s$^{-1}$. Also, steady flow in the range -10 km s$^{-1}$ to 10 km s$^{-1}$ results in four upward propagating wave modes whereas in the absence of it results in three wave modes only. When the strength of steady flow is kept between -10 km s$^{-1}$ to 10 km s$^{-1}$, the fast and thermal mode shows damping.

*3.2 Damping length behaviour as a function of the steady flow*

The behaviour of the wavelength, damping (or amplification) length, and damping per wavelength for various wave modes is shown in Figs. 1a, b, c. Figure 1a shows the wavelength of slow, fast and thermal modes as a function of steady flow ranging from -15 km s$^{-1}$ to 15 km s$^{-1}$. Figure 1b shows that the damping length of slow and fast mode shows minimum damping length in the absence of steady flow. When steady flow is present, the damping length increases for slow and fast modes. This effect is more pronounced in the case of slow mode. The damping length of thermal mode is maximum in the absence of steady flow. The damping length of thermal mode decreases in the presence of steady flow. Figure 1c shows the damping per wavelength for various wave modes. It is found that the fast mode shows strongest damping in the presence of steady flow. However, in the absence of steady flow thermal mode shows strongest damping. The damping length corresponding to various damping mechanisms is shown in Table 2 for the wave period of 5 min (where the photospheric power is maximum). It is evident that the thermal conduction does not play a role in damping except introducing an additional thermal mode and radiation is the most important damping mechanism. The damping length of magnetoacoustic waves is found to increase at the wave period of 5 min and a parallel steady flow of strength 5 km s$^{-1}$.

Figure 2 shows the behaviour of damping length for different values of prominence density. The damping length of slow, fast and thermal modes decreases with density. Figure 3 shows the influence of temperature on the damping length of wave modes. Figure 3a shows that the damping length of slow mode is minimum at temperature of 15000 K. In the absence of steady flow the damping length is minimum at temperature of 8000 K. Figure 3b shows that the damping length of fast mode is also minimum at temperature of 15000 K. The damping length of fast mode at the temperature of 15000 K is maximum in the absence of steady flow and it decreases further with flow. Figure 3c shows that the thermal mode damping length is minimum at temperature of 5000 K.

**Table 1.** Parameter values considered in the studied prominence regimes. All quantities are expressed in MKS units (after Carbonell et al. 2006).

| Regime | $T_0$ | $\rho_0$ | $\chi^*$ | $\alpha$ | $\tilde{\mu}$ | Reference |
|---|---|---|---|---|---|---|
| Prominence (1.1) | 8000 | $5 \times 10^{-11}$ | $1.76 \times 10^{-13}$ | 7.4 | 0.8 | Hildner (1974) |
| Prominence (1.2) | 8000 | $5 \times 10^{-11}$ | $1.76 \times 10^{-53}$ | 17.4 | 0.8 | Milne et al. (1979) |
| Prominence (1.3) | 8000 | $5 \times 10^{-11}$ | $1.76 \times 10^{-104}$ | 30 | 0.8 | Rosner et al. (1978) |

The influence of prominence magnetic field on the damping length of slow, fast and thermal mode is shown in Fig. 4a, b, c. The damping length of slow mode increases with magnetic field strength and this effect is more pronounced in the case of fast mode. The damping length of thermal mode shows change in behaviour only at the steady flow strength of 5 km s$^{-1}$ to 15 km s$^{-1}$. Figure 5 shows the influence of various prominence regimes on the damping length of wave modes. The different values of $\chi^*$ and $\alpha$ in various prominence regimes correspond to different radiative losses. Figure 5a shows that in the absence of steady flow, damping length of slow mode is minimum at prominence regime 1.1 and there is no change at higher values of steady flow that ranges between 5 km s$^{-1}$ to 15 km s$^{-1}$. Figure 5b shows that in the considered steady flow regime and prominence regime 1.1, the damping length of fast mode is found to minimum. Figure 5c shows that the damping length of thermal mode is influenced by various prominence regimes only when the anti-parallel steady flow is present. Figure 6 shows the influence of propagation angle on the damping length of slow, fast and thermal modes. Figure 6a shows that the damping length of slow mode remain unchanged with propagation angle. However, at the propagation angle of $\pi/2$ there is no slow mode present. Figure 6b shows that at the propagation angle of $10^{-2}$, the damping length of fast mode is maximum in the absence of steady flow and it further decreases with steady flow. Figure 6c shows that the thermal mode is absent at the propagation angle of π/2.

**Table 2.** Damping (or amplification) length $d_L$ (in m) for the case including flow (5 km s$^{-1}$), thermal conduction, radiation and heating (FTRH); thermal conduction, radiation and heating (TRH); only thermal conduction (TC); and only radiation (RAD) for a period of 300 s. The slow mode wave shows amplification whereas the fast and thermal mode wave shows damping.

|      | Thermal mode | Slow Mode | Fast Mode |
|------|---|---|---|
| FTRH | $1.2 \times 10^{-1}$ | $3.4 \times 10^{11}$ | $3.4 \times 10^{11}$ |
| TRH  | $9 \times 10^{2}$ | $2.3 \times 10^{6}$ | $1.4 \times 10^{10}$ |
| TC   | $1.5 \times 10^{3}$ | $1.7 \times 10^{11}$ | $1.4 \times 10^{11}$ |
| RAD  | No thermal mode | $2.3 \times 10^{6}$ | $1.5 \times 10^{10}$ |

## 4. Conclusions

In this paper, we have studied the effect of steady flow on the spatial damping of linear non-adiabatic magnetoacoustic waves by deriving a new general dispersion relation invoking the flow in MHD equations. The dispersion relation has been numerically solved to study the damping length of slow, fast and thermal modes.

The main conclusions of our study are;

1. The presence of steady flow breaks the symmetry of forward and backward propagating MHD wave modes in prominences.
2. The strength of steady flow has dramatic influence on the propagation and damping characteristics of magnetoacoustic and thermal mode.
3. We have explored the behaviour of damping (or amplification) length (= $1/k_I$) in the range of flow strength ranging between -15 km s$^{-1}$ to 15 km s$^{-1}$. For prominence regime 1.1, it is found that the damping length of slow mode increases with the strength of steady flow. The damping length of fast mode also shows the similar behavior. The damping length of thermal mode decreases with the strength of steady flow. This is a very important result because in the absence of steady flow all the forward propagating wave modes shows damping.
4. In the presence of steady flow, the fast mode wave shows strongest damping in comparison with slow and thermal modes. However, in the absence of steady flow the thermal mode is strongly damped.

5. The damping length of wave modes as a function of physical parameters has been explored. The damping length of magnetoacoustic and thermal mode depends depends strongly on the prominence density. It is found that the damping length of slow, fast and thermal modes decreases with density. The damping length of various modes depends on prominence temperature also. The damping length of slow and fast mode increases with magnetic field strength.

6. The influence of various prominence regimes that correspond to three different radiative losses on the damping length of wave modes has been studied. These radiative losses mimic both the optically thin and thick radiation. The damping length of slow mode is found to be minimum in the absence of steady flow and there is no appreciable change noted at the higher values. The damping length of thermal mode is influenced by various prominence regimes only when the anti-parallel steady flow is present.

7. The damping length of magnetoacoustic and thermal modes depends strongly on the propagation angle. At the propagation angle of $\pi/2$ the slow and thermal mode is found to be absent. At the propagation angle of $10^{-2}$, the damping length of fast mode is maximum in the absence of steady flow and it further decreases with steady flow.

8. The influence of various heating mechanisms on the damping length of wave modes is studied. It is found that the damping length of wave modes remain unchanged with the heating mechanisms.

It is now interesting to compare the theoretical damping length with the value derived from the observations. In Terradas et al. (2002), the dominant period of the oscillations, interpreted as slow modes, was around 75 min, and correspondingly the amplification length was found to be $2 \times 10^{13}$ m. In the absence of steady flow the slow mode at wave period of 75 min shows damping. For the wave period between 5 min and 15 min, the amplification length corresponding to slow mode, in the case of prominence regime 1.1 , varies between $3.4 \times 10^{11}$ m to $2 \times 10^{12}$ m.

In conclusion, taking account of the observations of steady flow for the *first time* in prominences we study the spatial damping of small-amplitude prominence oscillations. It has been found that the steady flow has dramatic influence on the propagation and damping of small-amplitude prominence oscillations and this may have potential importance for both wave detection and seismology in prominences.

*Acknowledgements*

KAPS acknowledges the Indian Institute of Astrophysics, Bangalore for the postdoctoral fellowship. We are deeply grateful to Ramon Oliver for his valuable suggestions on this work.

# FIGURES

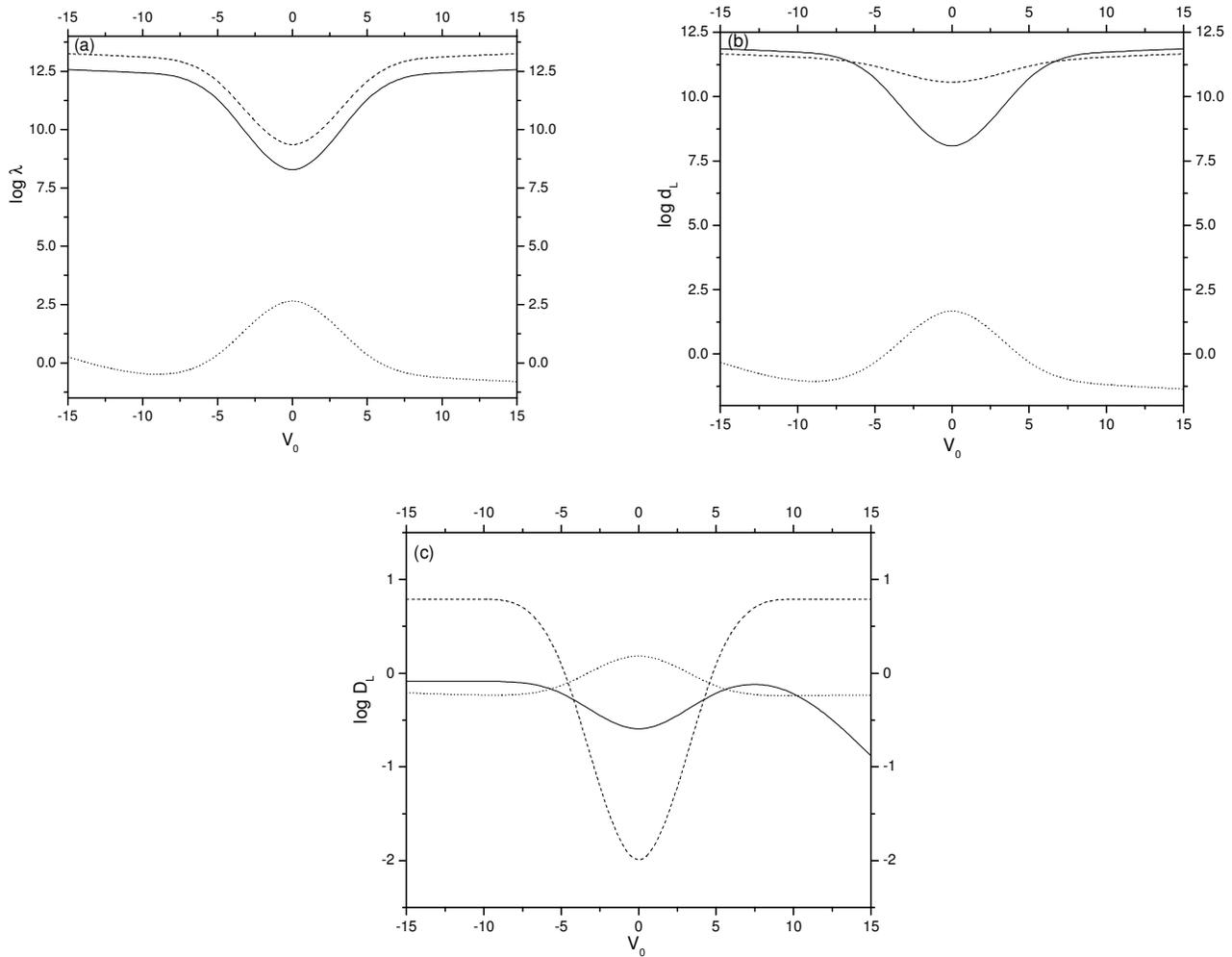

**Fig. 1.** (a) Wavelength = $2\pi/\mathrm{Re}(\omega)$, (b) damping length = $1/\mathrm{Im}(\omega)$ and, (c) damping per wavelength = $\mathrm{Im}(\omega)/\mathrm{Re}(\omega)$ as a function of steady flow for slow (solid line), fast (dashed line) and thermal mode (dotted line) and constant heating per unit volume ($a = b = 0$).

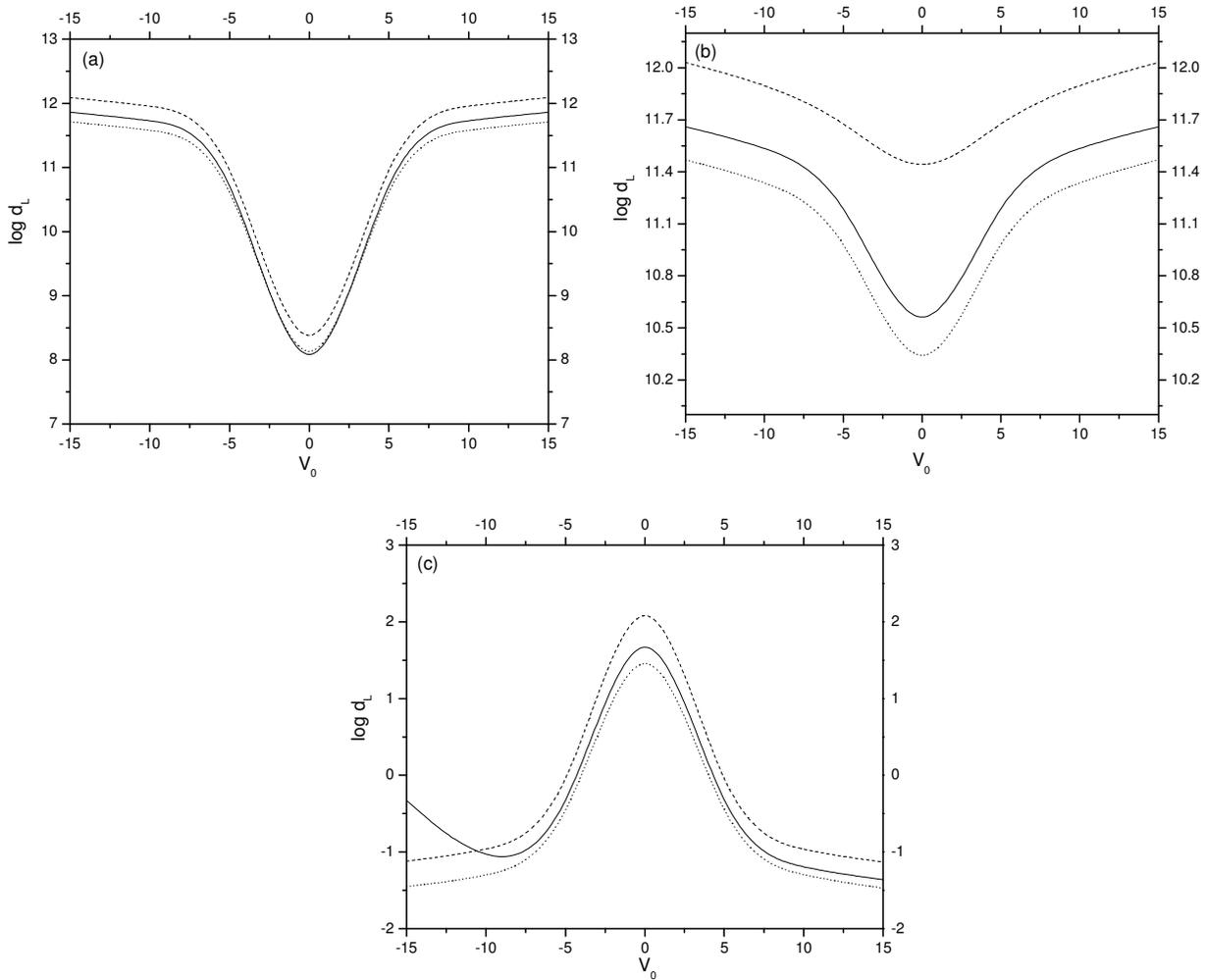

**Fig. 2.** Damping length as a function of steady flow for (a) slow, (b) fast and (c) thermal mode and constant heating per unit volume ($a = b = 0$). Solid line ($\rho_0 = 5\times10^{-11}$ kg m$^{-3}$); dashed line ($\rho_0 = 1\times10^{-11}$ kg m$^{-3}$); dotted line ($\rho_0 = 10\times10^{-11}$ kg m$^{-3}$).

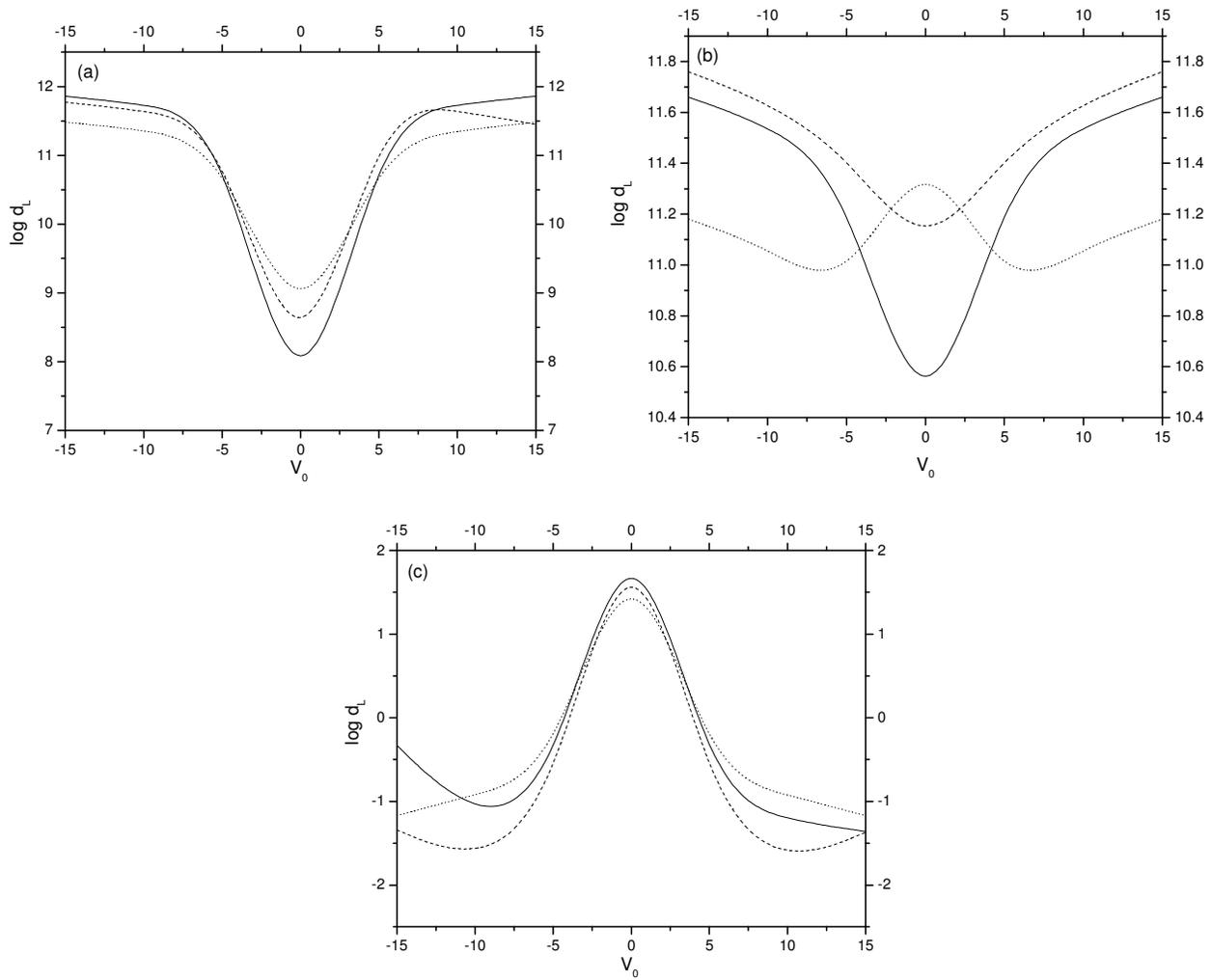

**Fig. 3.** Damping length as a function of steady flow for (a) slow, (b) fast and (c) thermal mode and constant heating per unit volume ($a = b = 0$). Solid line ($T_0 = 8000$ K); dashed line ($T_0 = 5000$ K); dotted line ($T_0 = 15000$ K).

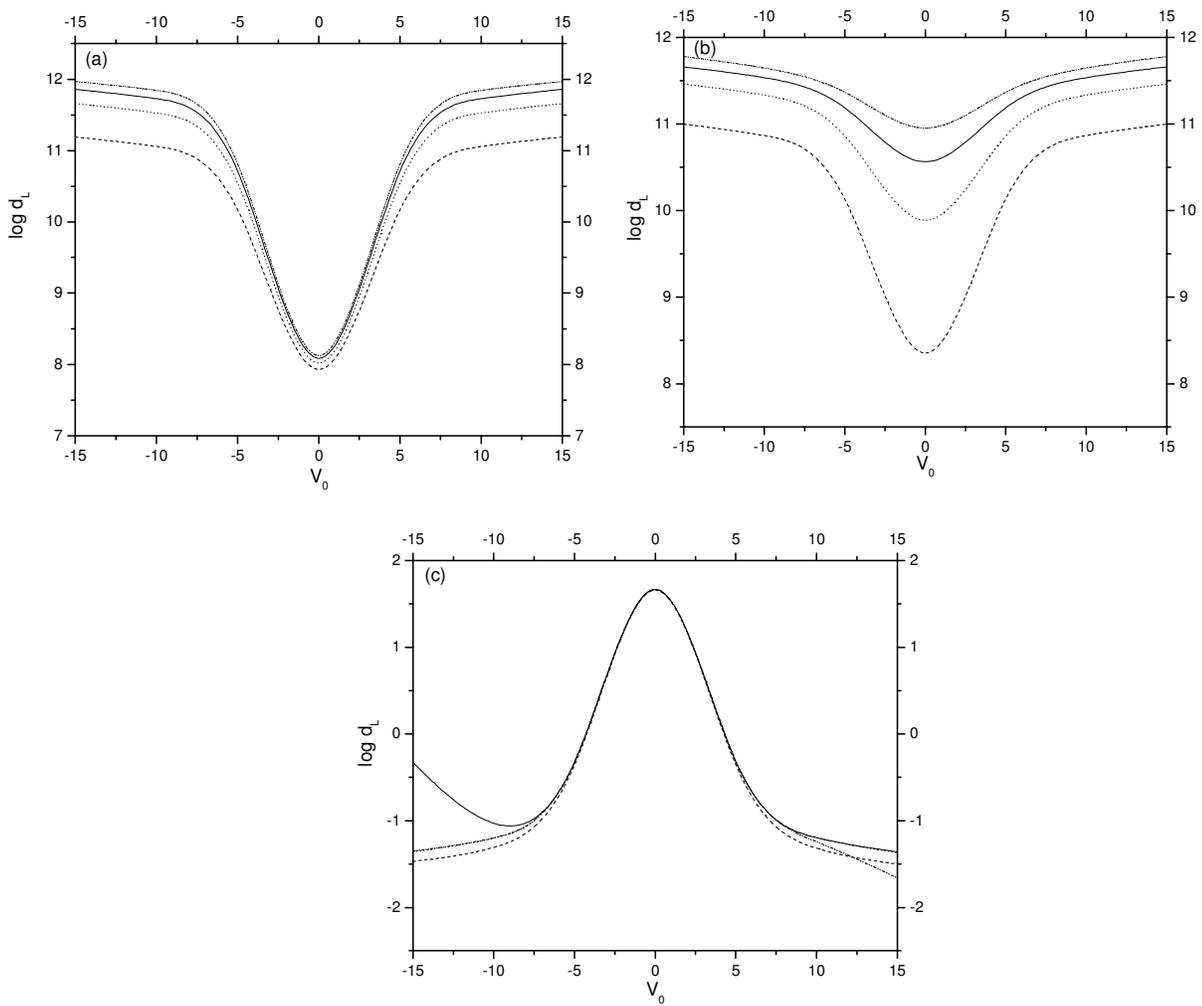

**Fig. 4.** Damping length as a function of steady flow for (a) slow, (b) fast and (c) thermal mode and constant heating per unit volume ($a = b = 0$). Solid line ($B_0 = 10\times10^{-4}$ T); dashed line ($B_0 = 1\times10^{-4}$ T); dotted line ($B_0 = 5\times10^{-4}$ T); dash dotted line ($15\times10^{-4}$ T).

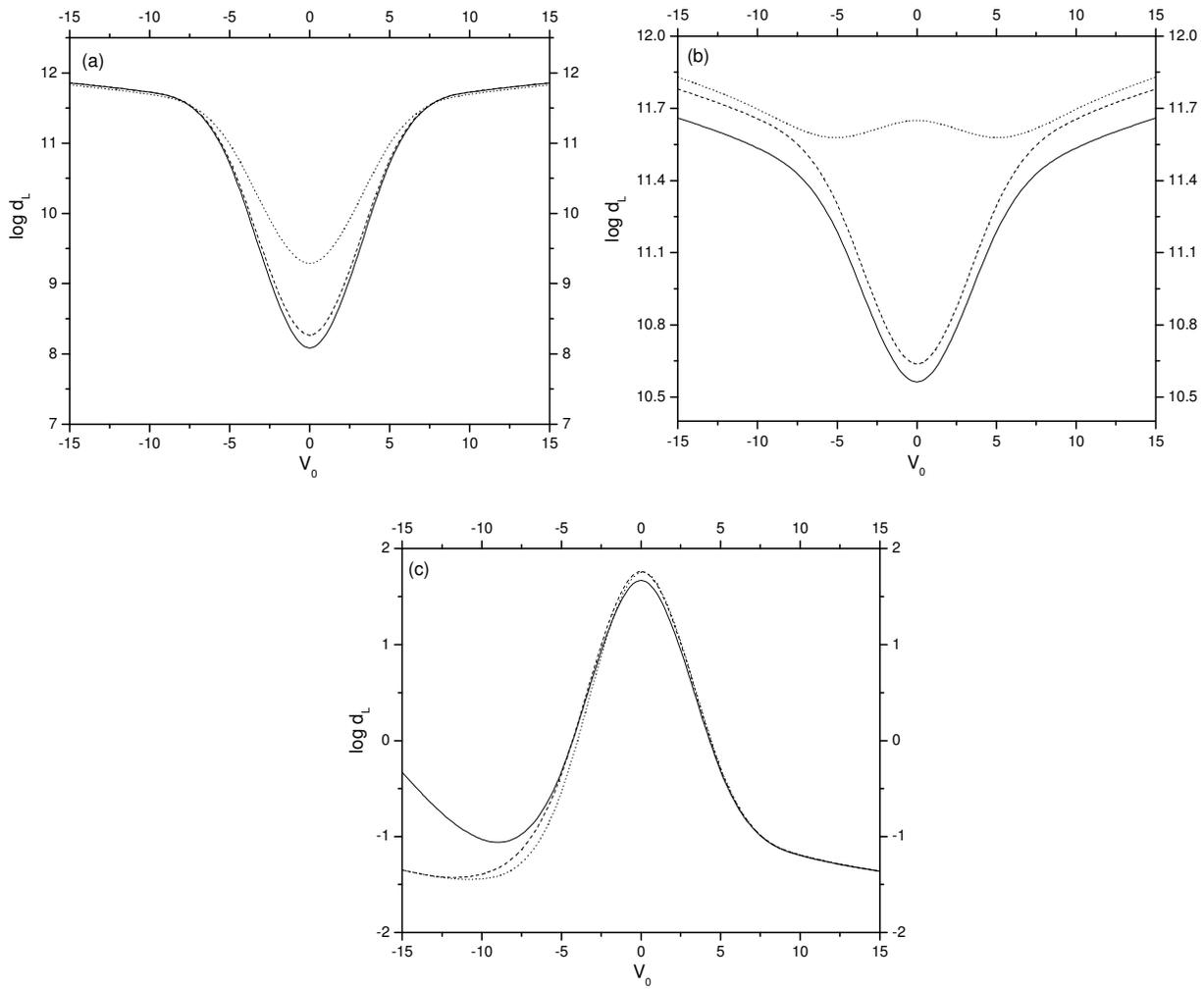

**Fig. 5.** Damping length as a function of steady flow for (a) slow, (b) fast and (c) thermal mode and constant heating per unit volume ($a = b = 0$). Solid line corresponds to prominence regime 1.1, dashed line corresponds to prominence regime 1.2 and dotted line corresponds to prominence regime 1.3.

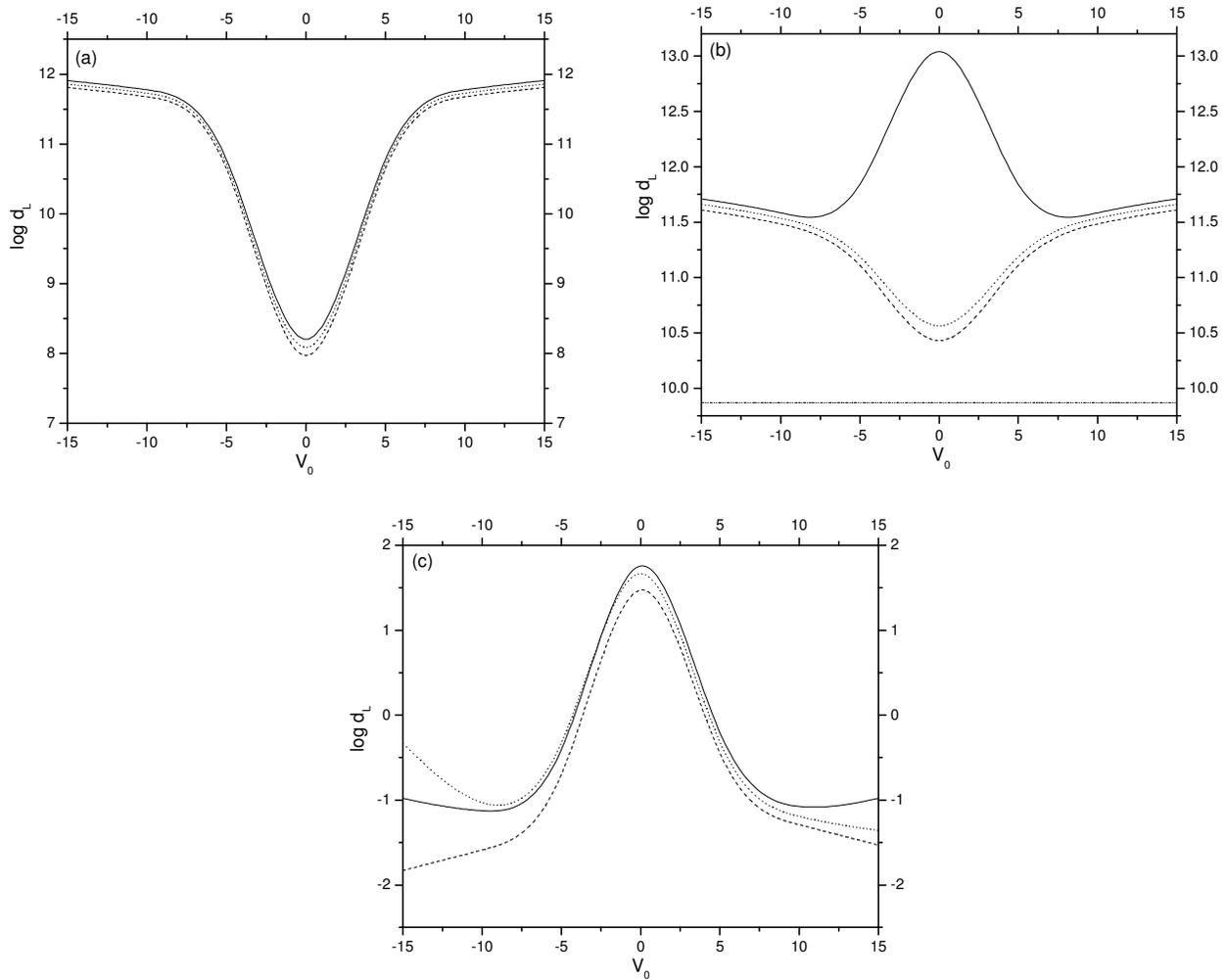

**Fig. 6.** Damping length as a function of steady flow for (a) slow, (b) fast and (c) thermal mode and constant heating per unit volume ($a = b = 0$). Solid line ($\theta = 10^{-2}$), dashed line ($\theta = \pi/3$), dotted line ($\theta = \pi/4$) and, dash-dotted line ($\theta = \pi/2$).